\documentclass{basi}
\usepackage{graphicx}
\usepackage{rotating}
\begin{document}
\title[Photometry of ASAS 134738$+$0410.1]{A CCD photometric study of the 
newly discovered contact binary ASAS 134738$+$0410.1}
\author[Deb et al.]%
       {Sukanta Deb$^1$ \thanks{e-mail: sdeb@physics.du.ac.in; 
sukantodeb@gmail.com}, Harinder P. Singh$^1$, T. R. Seshadri$^1$, 
Ranjan Gupta$^2$
       \\
 $^1$ Department of Physics \& Astrophysics, University of Delhi, Delhi 110007, India\\
 $^2$Inter-University Centre for Astronomy and Astrophysics (IUCAA), Post Bag 4, Ganeshkhind, Pune 411007, India}


\maketitle
\label{firstpage}
\begin{abstract}
We present a CCD photometric study of the star with ASAS ID 134738$+$0410.1 
using V band observations obtained from the $IUCAA$ Girawali Observatory (IGO) 2-metre telescope, India.~The star was selected from the $\delta$ Scuti 
database of All Sky Automated Survey (ASAS) (Pojmanski 2002). Our analysis 
reveals that the star is not a $\delta$ Scuti variable but is in fact a W UMa 
type contact binary with an orbital period of 0.2853067 day. Two new 
times of primary and secondary minima were determined from the observed data. 
A preliminary solution obtained using the Wilson-Devinney light curve 
modelling technique indicates that the star is more likely a 
partially-eclipsing W UMa type contact binary. However, the determination of 
actual subtype of this binary is quite impossible from the photometry alone, 
as the observed light curve can fitted for both A- and W-type solutions. The 
exact classification of this binary needs to be determined from high 
resolution spectroscopy.
\end{abstract}
\begin{keywords}
binaries: close - stars : individual ASAS 134738$+$0410.1 - stars: magnetic 
fields
\end{keywords}
\section{Introduction}
\label{sec:intro}
In the ASAS\footnote{http://www.astrouw.edu.pl/asas/} database, the star has
maximum V magnitude of 13.47 mag and is located at $\alpha = 13^{\rm h}\,
47^{\rm m}\,38^{\rm s}.0$ and $\delta = 04^{\rm o}\,10^{\rm '}\,05^{\rm ''}.9$ 
(Epoch J2000). The star has been classified as $\delta$ Scuti/Contact/
Semi-Detached binary with a period of P$_{\rm ASAS}$ = 0.142654 day. The 
classification is thus ambiguous. In order to know true nature of this object, we have obtained high precision CCD photometric data from new observations. 
The new data indicates that the star is a W UMa contact binary with an orbital 
period of P$_{\rm IGO} = 0.2853067$ day which is nearly twice the period 
P$_{\rm ASAS}$ = 0.142654 day.  
 
\begin{table*}
\begin{center}
\caption{Basic information of ASAS 134738$+$0410.1, the comparison star and the 
check star}
\label{Table1}
\scalebox{0.82}{
\begin{tabular}{lccccc}
\hline
Star  & RA (J2000) & DEC (J2000) & J [mag] & H [mag]& K [mag]  \\ \hline
ASAS 134738$+$0410.1 (Target) & $13^{\rm h}\,47^{\rm m}\,38^{\rm s}.00$ & $04^{\rm o}\,10^{\rm '}\, 06^{\rm ''}.00$ &11.894 &11.472 &11.407  \\

2MASS 13473475$+$0408167 (Comparison) & $13^{\rm h}\,47^{\rm m}\,34^{\rm s}.76$ & $04^{\rm o}\,08^{\rm '} 16^{\rm ''}.72$ &12.408&12.026 &11.951\\ 
2MASS 13472306$+$0413420 (Check) & $13^{\rm h}\,47^{\rm m}\,23^{\rm s}.06$ & $04^{\rm o}\,13^{\rm '}\,42^{\rm ''}.10$ &12.213 &11.870 &11.789 \\ \hline
\end{tabular}}

\end{center}
\end{table*}

The V band CCD photometric observations of the star were carried out 
with the IGO 2-m telescope, located about 80 km from Pune, India during two 
nights on March 31 and April $02$  in 2009.~The IUCAA Faint Object 
Spectrograph Camera (IFOSC) equipped with EEV 2 K$\times$ 2 K thinned, 
back-illuminated CCD with 13.5 $\mu m$ pixels was used. The CCD used for 
imaging provides an effective field of view of $\sim 10.5^{'} 
\times 10.5^{'} $ on the sky corresponding to a plate scale of 0.3 arcsec 
pixel$^{-1}$. The gain and read out noise of the CCD camera are 1.5 e$^{-}$/ADU
 and 4 e$^{-}$ respectively. The FWHM of the stellar image varied from 3 to 
5 pixels during the observations. We took a total of 153 frames in the V band 
with the exposure times varied between 100 s and 180 s for a good photometric 
accuracy. 

The co-ordinates of the variable, comparison star and the check star along 
with the infrared JHK magnitudes taken from the 2MASS catalogue (Cutri et al. 
2003) are listed in Table 1. The comparison and the check star are so close to 
the variable that they are in the same field during the observations. 
Image pre-processing and data reduction was carried out using IRAF\footnote 
{IRAF is distributed by the National Optical Astronomy Observatories, which 
are operated by the Association of Universities for Research in Astronomy, 
Inc., under cooperative agreement with the National Science Foundation.} and 
MIDAS softwares. Instrumental magnitudes were obtained using the DAOPHOT 
package (Stetson 1987, 1992). The various tasks, e.g., $find, phot, daogrow, 
daomatch$ and $daomaster$ were applied in order to obtain the instrumental 
magnitudes of stars in all the frames. Extinction corrections were ignored as 
the target star is very close to the comparison star. In Fig.~1, we show the plots of the differential V band magnitude of (Variable - Comparison), 
(Comparison - Check) versus Heliocentric Julian Day (HJD) in the left, right 
upper and lower panels respectively for observations on March 31 and April 02, 
2009. The reduced results show that the difference between the magnitude of 
the check star and that of the comparison star was constant with a probable 
error of $\pm$0.004 mag in the V band \footnote{The observational data 
presented in this paper can be obtained from the authors on request.}.
\section{Period analysis and determination of times of minima}
A period analysis was carried out using multi-harmonic ANOVA algorithm 
technique developed by Schwarzenberg-Czerny (1996, hereafter SC96)  to find out 
the period of ASAS 134738$+$0410.1.~The method computes periodogram by fitting 
multi-harmonic Fourier series to the time series data. It is very 
efficient and robust for non-sinusoidal signals. The various statistical 
properties and advantages of this method are described in SC96. The following 
times of minima were determined from the data using the Kwee \& van Woerden's 
method (1956).
\begin{displaymath}
\rm Min~I~[\rm Primary~\rm Minimum] = 2454922.27125(5), 2454924.26926(28) 
\end{displaymath}
 
\begin{displaymath}
\rm Min~II~[\rm Secondary~\rm Minimum] = 2454922.41364(5), 2454924.41099(8) 
\end{displaymath}
The numbers given in parentheses represent the probable errors and are expressed
in terms of the last quoted digits. For example, 2454922.27125(5) should 
be interpreted as $2454922.27125 \pm 0.00005$. We use the following ephemeris 
to derive the phased light curve.
\begin{eqnarray}
\rm Min\,I = \rm HJD\,2454922.27125(5) + 0^{d}.2853076 \times E\,, 
\end{eqnarray}
where E is the epoch in days. We plot the phased light curve of the star in 
Fig.~2. The phased light curve of the star is defined as
an array of phase ($\Phi$) and differential V band magnitude ($\rm \Delta V$).
The term phase ($\Phi$) is defined as :
\begin{equation}
\Phi =\frac{\left( t-t_{0}\right) }{P}-{\rm Int}\left( \frac{ t-t_{0}
}{P}\right).
\end{equation}
The value of $\Phi$ is from 0 to 1, corresponding to a full cycle of
the orbital period ($P$) and $\rm Int$ denotes the integer part of the 
quantity. The zero point of the phase corresponds to the time of primary 
eclipse ($t_{0}$).
\begin{figure}
\begin{center}
\includegraphics[height=8cm,width=9cm]{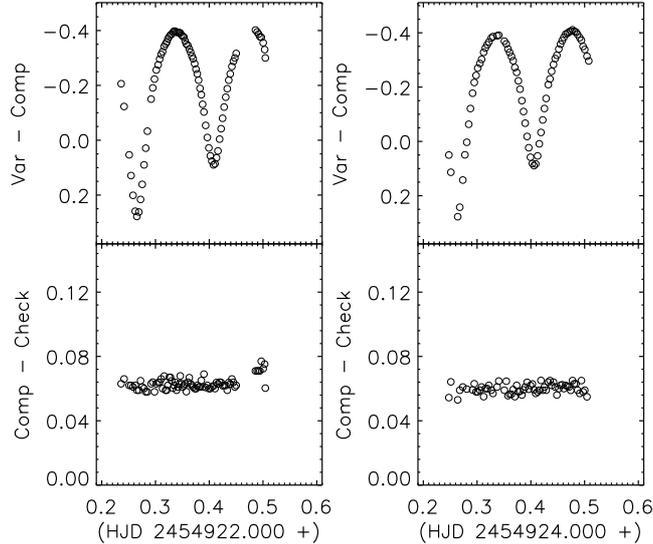}
\caption{Left upper and lower panels show the differential magnitude versus 
time (in HJD) of the variable star (Var) with respect to the comparison star 
(Comp) and comparison star with respect to the check star (Check) respectively 
for March 31, 2009. The right upper and lower panels show the corresponding 
plots for April 02, 2009.}
\end{center}
\end{figure}           
\begin{figure}
\begin{center}
\includegraphics[height=8cm,width=9cm]{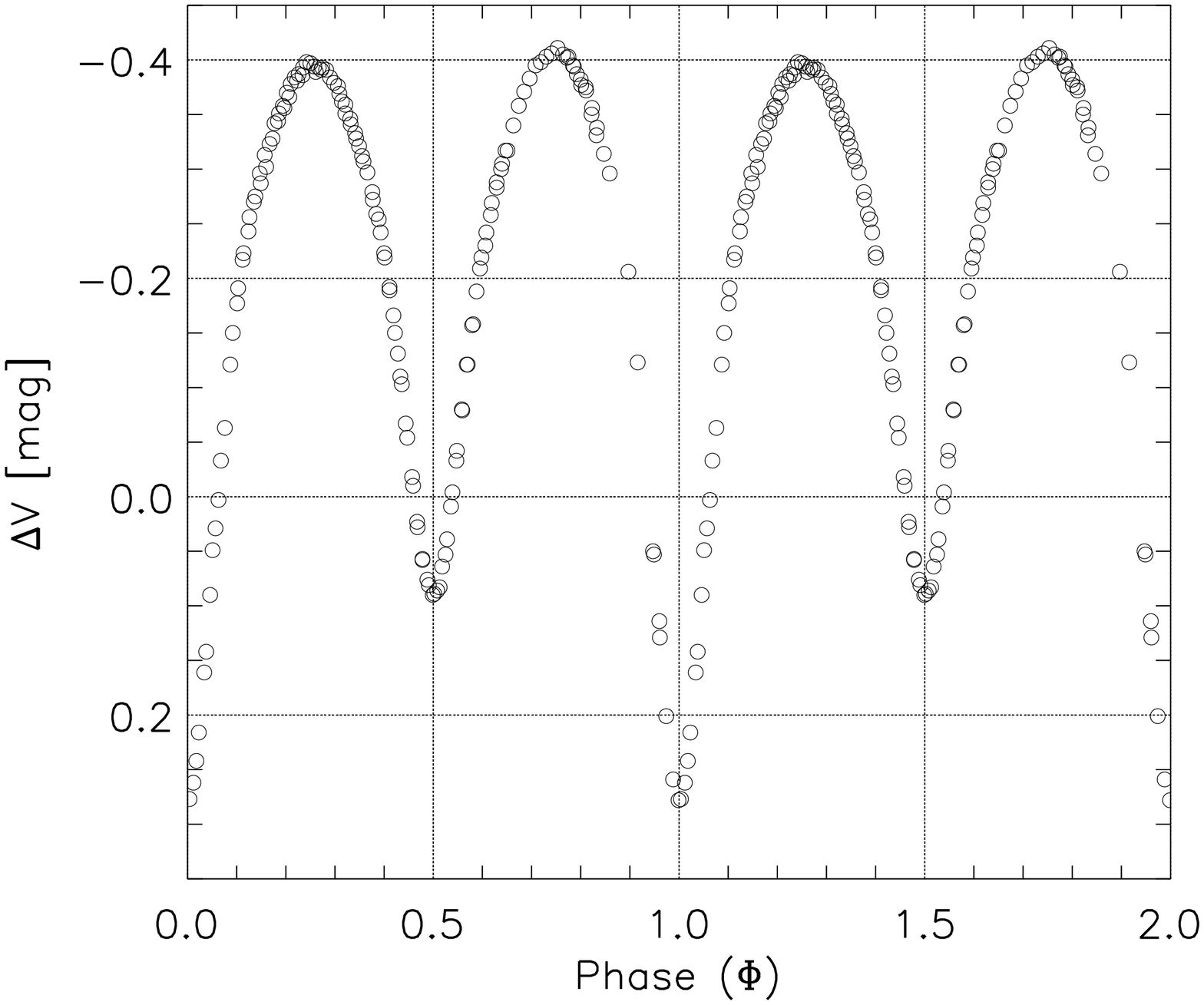}
\caption{Phased light curve of ASAS 134738$+$0410.1. The light 
curve is plotted in the range [0, 2] for a clear visualisation of the primary 
and the secondary minimum. Open circles denote the observational data points.}
\end{center}
\end{figure} 

A closer look at the light curve of ASAS 134738$+$0410.1 (Fig.~2) shows that the light curve exhibits O'Connell effect (O'Connell 1951), a phenomenon common 
to many W UMa type contact binaries due to the surface magnetic activities.
To quantify the O'Connell effect, we do a parabolic least-square fitting 
around the two maxima. The result indicates that the phase at 0.75 (Max II) is 
$\sim 0.008$ mag brighter than  that at phase 0.25 (Max I). Also from the 
light curve, it is clearly seen that the Min\,I and Min\,II have different 
eclipse depths, which indicates that the effective temperatures of the two 
stars are different. Therefore, the star is a W UMa type eclipsing binary not 
in thermal contact. 

The analysis of the light curve was performed by the 
software package PHOEBE (Pr\v{s}a \& Zitter 2005). It is a modified package of 
the widely used Wilson-Devinney (hereafter, WD) program for eclipsing binary 
stars (Wilson \& Devinney 1971; Wilson 1979; Wilson 1990).

The effective temperature of ASAS 134738$+$0410.1 can be calculated from the 
period-color relation given by Rucinski (2000) who derived the following 
relation for contact binary systems.
\begin{eqnarray}
(B-V) = 0.04\,P^{-2.25},
\end{eqnarray}  

where $P$ is the period in days.~With the above equation, the color of 
 ASAS 134738$+$0410.1 can be calculated as: $(B-V)$ = 0.672. The interstellar 
extinction along the direction of the star is $E(B-V) = 0.023$ following 
Schlegel et al. (1998). Therefore, the intrinsic color index of the star would 
be $(B-V)_{0} = 0.649$. On the other hand, the infrared color index for the 
star is $(J-K) = 0.487$ following Cutri et al. (2003). Both these color 
indices suggest a spectral type nearly $\rm G{5}V$ for the binary system.

The adopted parameters are the temperature of the star 1 $T_{1}$ = 5560 K 
suitable for the spectral type G5V through the calibration of Cox (2000), the 
limb-darkening coefficients $x_{1} = x_{2}$, $y_{1} = y_{2}$ interpolated for 
square root law from Van Hamme (1993) tables, coefficients of gravity 
darkening $g_{1}$ = $g_{2} = 0.32$ (Lucy 1967), the bolometric albedos
$A_{1}$ = $A_{2} = 0.5$ (Rucinski 1969) appropriate for convective envelopes. 

The adjustable parameters are the orbital inclination $i$, mass ratio $q_{ph}$, 
the mean temperature of the star 2 $T_{2}$, the surface potentials of the 
components $\Omega_{1}$ and  $\Omega_{2}$, the monochromatic luminosity of the 
star 1 $L_{1}$. Planck function was used to compute the luminosity of the star 
1. 

Since there is no spectroscopic mass ratio available presently, we try to 
find the photometric mass ratio $q_{ph}$ using trial values in the range
0.3 to $3.80$ in steps of 0.1. Assuming that initially the system is a detached system, the differential  corrections were started from mode 2, the 
differential correction converged to a mode 3 solution (contact mode).
We tried both A- and W-type solutions for the system. Both the solutions 
gave plausible fit to the observed light curve over a wide range of $q_{ph}$ 
values due to its partial eclipsing nature. Therefore, it is impossible to 
reliably determine its actual subtype and other parameters from the 
photometric light curve alone, unless high resolution spectroscopic 
observations are obtained. 
\section{Summary and conclusions}
We have discovered that star having ASAS ID 134738+0410.1 identified by ASAS 
is a W UMa type overcontact binary.  We have determined new period and times 
of minima from very accurate and precise CCD data. Based on the WD code as 
implemented in the software PHOEBE, we have done preliminary modelling of the 
light curve. However, unique value of the mass ratio could not be obtained
from the modelling, due to its partial-eclipsing nature. Hence, the actual 
subtype of the system could not be determined. This star deserves high 
resolution spectroscopic time series radial velocity measurements to determine 
its true subtype and mass ratio which should be combined with the photometric 
light curve data to obtain various geometrical and physical parameters 
accurately.
\section{Acknowledgement}
The authors thank $IUCAA$ for providing telescope time available on the IGO 2-m 
telescope. The authors thank the anonymous referee for many useful
comments and suggestions which improved the presentation of the paper. The 
authors thank Dr.~Vijay Mohan for helpful discussions. SD thanks CSIR, Govt. 
of India, for a Senior Research Fellowship. The use of the SIMBAD, ADS, ESO 
DSS databases is gratefully acknowledged. This publication makes use of data 
products from the Two Micron All Sky Survey, which is a joint project of the 
University of Massachusetts and the Infrared Processing and Analysis 
Center/California Institute of Technology, funded by the National Aeronautics 
and Space Administration and the National Science Foundation. ESO-MIDAS
was used as a part of the data analysis.

\label{lastpage}

\end{document}